\documentclass{emulateapj}

\newcommand{\chandra}{{\it Chandra}}
\newcommand{\swift}{{\it Swift}}

\newcommand{\suzaku}{{\it Suzaku}}
\newcommand{\NOBS}{31}
\newcommand{\NSSS}{twelve}
\newcommand{\dend}{390}

\begin{document}

\markboth{J.-U. Ness et al.}{\swift\ monitoring of V458\,Vul}

\title{\swift\ X-ray and UV monitoring of the Classical Nova V458\,Vul (Nova Vul 2007)}

\author{J.-U. Ness\altaffilmark{1,2}, J.J. Drake\altaffilmark{3},
  A.P. Beardmore\altaffilmark{4},
  D. Boyd\altaffilmark{5},
  M.F. Bode\altaffilmark{6},  
  S. Brady\altaffilmark{7},
  P.A. Evans\altaffilmark{4},
  B.T. Gaensicke\altaffilmark{8},
  S. Kitamoto\altaffilmark{9}, 
  C. Knigge\altaffilmark{10},
  I. Miller\altaffilmark{11},
  J.P. Osborne\altaffilmark{4}, 
 K.L. Page\altaffilmark{4},
  P. Rodriguez-Gil\altaffilmark{12,13},
 G. Schwarz\altaffilmark{14},
  B. Staels\altaffilmark{15},
D. Steeghs\altaffilmark{8,3},
 D. Takei\altaffilmark{9},
  M. Tsujimoto\altaffilmark{16,17},
R. Wesson\altaffilmark{18},
 A. Zijlstra\altaffilmark{19}
}

\altaffiltext{1}{European Space Astronomy Centre, PO Box 78, 28691 Villanueva de la Ca\~nada, Madrid, Spain: juness@sciops.esa.int}
\altaffiltext{2}{School of Earth and Space Exploration, Arizona
State University, Tempe, AZ 85287-1404, USA}
\altaffiltext{3}{Harvard-Smithsonian Center for Astrophysics, 60
  Garden Street, Cambridge, MA 02138, USA}
\altaffiltext{4}{Department of Physics \& Astronomy, University of
  Leicester, Leicester, LE1 7RH, UK}
\altaffiltext{5}{BAA\,VSS, 5 Silver Lane, West Challow, Wantage, OX12 9TX, UK}
\altaffiltext{6}{Astrophysics Research Institute, Liverpool John
  Moores University, Twelve Quays House, Egerton Wharf, Birkenhead
  CH41 1LD, UK}
\altaffiltext{7}{AAVSO, 5 Melba Drive, Hudson, NH 03051, USA}
\altaffiltext{8}{Department of Physics, University of Warwick,
  Coventry CV4 7AL, UK}
\altaffiltext{9}{Department of Physics, Rikkyo University, 3-34-1
  Nishi-Ikebukuro, Toshima, Tokyo 171-8501, Japan}
\altaffiltext{10}{School of Physics \& Astronomy, University of
  Southampton, Southampton SO17~1BJ, UK}
\altaffiltext{11}{BAA\,VSS, Furzehill House, Ilston, Swansea, SA2 7LE, UK}
\altaffiltext{12}{Isaac Newton Group, PO Ap.\ de Correos 321, 38700
  Sta.\ Cruz de la Palma, Spain}
\altaffiltext{13}{Instituto de Astrof\'\i sica de Canarias, V\'\i a L\'actea s/n, La Laguna, E-38205 Santa Cruz de Tenerife, Spain}
\altaffiltext{14}{Department of Geology and Astronomy, West Chester
  University, West Chester, PA~19383, USA}
\altaffiltext{15}{CBA Flanders, Alan Guth Observatory, Koningshofbaan 51, Hofstade, Aalst, Belgium}
\altaffiltext{16}{Department of Astronomy \& Astrophysics, Pennsylvania State University,\\
  525 Davey Laboratory, University Park, PA 16802, USA}
\altaffiltext{17}{Japan Aerospace Exploration Agency, Institute of
  Space and Astronautical Science, 3-1-1 Yoshino-dai, Sagamihara,
  Kanagawa 229-0015, Japan}
\altaffiltext{18}{Department of Physics \& Astronomy, University
  College London, Gower Street, London WC1E 6BT, UK}
\altaffiltext{19}{Jodrell Bank Center for Astrophysics,
Alan Turing Building, School of Physics and Astronomy,
The University of Manchester, Oxford Street, Manchester,
M13 9PL, UK}

\begin{abstract}
We describe the highly variable X-ray and UV emission of V458\,Vul (Nova 
Vul 2007), observed by \swift\ between 1 and 422 days after outburst. 
Initially bright only in the UV, V458\,Vul became a variable hard X-ray 
source due to optically thin thermal emission at k$T=0.64$\,keV with an X-ray band 
unabsorbed luminosity of $2.3\times 10^{34}$\,erg\,s$^{-1}$ during days 71-140.
The X-ray spectrum at this time requires a low Fe abundance
($0.2^{+0.3}_{-0.1}$ solar), consistent with a \suzaku\ measurement around
the same time. On day 315 we find a new X-ray spectral component which can
be described by a blackbody with temperature of k$T=23^{+9}_{-5}$\,eV,
while the previous hard X-ray component has declined by a factor of 3.8.
The spectrum of this soft X-ray component resembles those typically seen in
the class of supersoft sources (SSS) which suggests that the nova ejecta were
starting to clear and/or that the WD photosphere is
shrinking to the point at which its thermal emission reaches into the
X-ray band. We find a high degree of variability in the
soft component with a flare rising by an order of magnitude in count rate
in 0.2 days. In the following observations on days 342.4-383.6, the soft
component was not seen, only to emerge again on day 397. The hard component
continued to evolve, and we found an anticorrelation between the
hard X-ray emission and the UV emission, yielding a Spearman rank
probability of 97\%. After day 397, the hard component was still present,
was variable,
and continued to fade at an extremely slow rate but could not be analysed
owing to pile up contamination from the bright SSS component.
\end{abstract}


\section{Introduction}

Classical Novae (CNe) are caused by nuclear explosions in cataclysmic
variables. A history of accretion of hydrogen-rich material from a
main-sequence star onto the white dwarf (WD) primary can lead to
sufficient pressure in the surface envelope for thermonuclear runaway
to occur. Enough radiation is produced to launch a radiatively-driven
wind that ejects the outer envelope and forms an optically-thick shell
around the WD within which the high-energy radiation that is produced
by nuclear burning is initially trapped. The radiative output of the
nova then occurs primarily in the optical, but as the mass ejection
rate decreases, the radius of approximately unit
opacity shrinks, and successively hotter layers become visible
\cite[see, e.g.,][]{gallstar78}. If
nuclear burning continues long enough, the peak of the spectral energy
distribution eventually shifts into the X-ray regime, and at that time
the nova emits an X-ray spectrum that resembles those typically
observed in the class of supersoft X-ray sources
\citep[SSS:][]{kahab}. For a review of nova evolution see, e.g.,
\cite{st08}.

Early X-ray emission has been detected in several cases that was much
harder than typical SSS spectra
\citep[e.g.,][]{lloyd92,krautt96,mukai01,Orio2001}. The origin of this
emission is not entirely clear because it has never been observed with
sufficient signal-to-noise ratio and spectral resolution to constrain
theoretical models. The observed spectra are generally a good match
to optically thin plasma models
\citep[e.g.,][]{krautt96,balm98,Orio2001}, which suggests that the
plasma is collisional. \cite{mukai01} argued that shocks within
the nova ejecta were responsible, and this is consistent with
earlier work by \cite{lloyd92} and \cite{obrien94}. In some cases,
hard emission line spectra have also been observed after the nova had
turned off \citep[e.g., V382\,Vel:][]{ness_vel}. This has been
attributed to nebular emission that originates from the surrounding
medium that had been heated by radiation from the nova. However, some
of the emission lines that have been observed arise at energies too
high to be explained by photoexcitation. It is possible that these
lines are part of the continuing hard component, which would imply
that the cooling time scales of the hard component are extremely
long, up to years.

A different situation is the case of
symbiotic novae such as RS\,Oph, where the secondary is a giant with a
dense stellar wind. The ejecta run into the stellar wind, dissipating some
kinetic energy, producing X-ray radiation. The luminosity of the
emission from these shock systems is much higher than observed in Classical
Novae, and well-exposed X-ray spectra with high spectral resolution
have been obtained \citep{rsophshock}. In Classical Novae, the
secondary is a Main-Sequence star where the wind is not dense enough
for this scenario \citep{obrien94}.\\

In order to understand the evolution of the X-ray emission of novae,
more systematic observations are required, as exemplified by recent
\swift\ campaigns on RS\,Oph and V2491\,Cyg (\citealt{osborne06},
\citealt{bode06}, Page et al., in preparation).
Here, we present \swift\ monitoring observations of the nova
V458\,Vul, which was discovered by H. Abe 2007, August 8.54 at 9.5 mag
(\citealt{v458discovery}; see also \citealt{IBVS5803}).
\cite{buil07} found P-Cygni profiles in H-$\alpha$, $\beta$, and
$\gamma$ lines, as well as in various He\,{\sc i} lines with
full-width at half maximum (FWHM) corresponding to $\sim
1700-1900$\,km\,s$^{-1}$. On these grounds, they classified it as a
Classical Nova. Images obtained by the {\it IPHAS} survey
\citep{Drewetal05} shortly
before the nova outburst, reveal a point source nova progenitor
surrounded by a faint shell. Follow-up imaging and spectroscopy
imply this is a slow moving and massive shell of a planetary nebula
(PN) as opposed to a fast-moving and much lower mass nova remnant
\citep{Wessonetal08}. The only other likely example of a Classical Nova 
occurring inside a PN was Nova GK\,Per 1901 \citep{bode04}. The
optical decline of V458\,Vul was observed in detail by
\cite{poggiani08}, and the time scale $t_2$ (time for decline by 2
mag) was estimated at 7 days. The maximum magnitude versus rate of
decline relationship (MMRD) for the smoothed light curve implies a
distance
of 6.7-10.3\,kpc \citep{poggiani08}, and the relation derived by
\cite{DVL95} leads to a distance of $8.8\,\pm\,0.8$\,kpc.
However, the optical light curve showed two
strong flare-like peaks on days 4 and 10 after outburst
\citep{tarasova07}. This complicates the determination of a
reliable decline time, and the MMRD relationship method may not
work very well for novae with such erratic early light curves.
\citet{Wessonetal08}
found a distance of 13\,kpc based on light travel time from the
nova to the flash-ionized nebula and other methods,
and we adopt this value for our purposes.
The interstellar extinction ($A_{V,\rm{ISM}}$) was estimated to be
$1.76\,\pm\,0.32$\,mag \citep{poggiani08}. This converts to an
interstellar hydrogen-equivalent extinction column density of
$N_{\rm H,ISM}= (3.15\,\pm\,0.57)\times 10^{21}$\,cm$^{-2}$
using the relation $N_{\rm{H,ISM}}$/$A_{V,\rm{ISM}}= 1.79 \times
10^{21}$\,cm$^{-2}$\,mag$^{-1}$ \citep{predehl95}. 
The HEASARC $N_{\rm H}$ tool\footnote{http://heasarc.gsfc.nasa.gov/cgi-bin/Tools/w3nh/w3nh.pl}
calculates the total Galactic H\,{\sc i} column density for any
direction using the Leiden/Argentine/Bonn \citep[LAB; ][]{Kal05}
and \citet{dickey90} Galactic \ion{H}{1} surveys. For a cone of
radius $0.5\arcdeg$ centered on the J(2000) coordinates of
V458\,Vul, the LAB and \citet{dickey90} maps give an average
$N_{\rm H}$ value of $3.7\times10^{21}$\,cm$^{-2}$.
\cite{2007IAUC.8883....1L} found $E(B-V)=0.6$, which converts
to $3.6\,\pm\,1.2\times10^{21}$\,cm$^{-2}$ using the relation
$N_{\rm H}/E(B-V)=6\,\pm\,2\times10^{21}$\,cm$^{-2}$ by
\cite{dickey90,bohlin78}.

 A number of photometric periods were reported by
\cite{v458orbit}, and they attributed a periodicity of
0.58946\,days (14.147 hours) to the orbital modulation.
We caution, however, that different periods have
been detected in different observations \citep{Wessonetal08}
and finding the orbital period is not a straightforward task.
 Apart from the two flares in the early optical light curve,
\cite{tarasova07} found changes in the line widths that were
related to these brightenings. Such events have been observed in
other novae as well (e.g., in V723\,Cas or V2362\,Cyg,
\citealt{goranskij07} and \citealt{lynch08}, respectively).
The nature of rebrightenings in novae is unclear, and secondary
ejection events or interactions with the accretion disk have been
proposed \citep{lynch08}.

The first X-ray observation of V458\,Vul was obtained 1.2 days
after outburst\footnote{All times are given in days after discovery,
2007, August 8.54, and are mid-times between start and end of
the observations} with the X-Ray Telescope (XRT) aboard \swift\
but yielded no detection (see below).
Observations at this early stage of nova evolution are justified
by the X-ray detection of V838\,Her approximately 5 days after
discovery \citep{lloyd92}. 70 days later, another \swift\
observation of V458\,Vul was carried out, and \cite{v458atel}
found 192 X-ray counts in a 6.7-ks exposure. From thermal models
they estimated a plasma temperature of k$T=0.2^{+0.40}_{-0.07}$\,keV
and an absorbing column of
$N_{\rm H}=8^{+2}_{-3}\times10^{21}$\,cm$^{-2}$. Around day 88
after outburst (2007, November 4), a \suzaku\ observation with
medium spectral resolution and higher $S/N$ than obtained by
\swift\ was carried out using the X-ray Imaging Spectrometer
\citep[XIS:][]{koyama07}, and \cite{masahiro08} found a plasma
temperature of $0.64\,\pm\,0.07$\,keV and
$N_{\rm H}=3.1^{+1.8}_{-1.3}\times10^{21}$\,cm$^{-2}$ from
optically-thin collision-dominated plasma emission models. Since the
\suzaku\ detectors have higher spectral resolution than the \swift\ XRT,
individual lines can be identified, and the spectrum is clearly an
emission line spectrum. \cite{masahiro08} investigated the chemical
composition and found overabundances of Ne, Mg, Si and S, and an
underabundance of Fe. For O they determined an upper limit of 1.5
times solar. On day 397, the nova entered the SSS phase with a
highly variable \swift\ XRT count rate \citep{v458sss}. The
aim of this paper is the study of the evolution of the hard
X-ray component that was first observed on day 71 after outburst
\citep{v458atel}. In \S\ref{obssect} we describe four observing
campaigns with optical, ultraviolet, and X-ray observations.
For the \swift\ observations taken after day 397 we only analyze
the hard X-ray emission component and postpone the analysis of
the SSS phase to a later paper that will be published after
the SSS phase has ended. In \S\ref{disc} we discuss our results,
and summarize our conclusions in \S\ref{concl}.

\section{Observations}
\label{obssect}

 In this paper we present \NOBS\ \swift\ XRT \citep{xrt} and
UVOT \citep{uvot} observations taken between August 2007 and
September 2008. The observations
are grouped in four campaigns (see Table~\ref{tab1}), one observation
only 1.2 days
after the outburst, ten observations taken between days 71 and
140.5, eight observations taken between days 315 and 390, and
\NSSS\ observations taken during the SSS phase, after day 390
\citep{v458sss}. The fourth campaign was a high-density campaign
in response to the emergence of the bright SSS component. In this
paper we analyze only the hard emission component of the observations
taken during the fourth campaign and postpone the analysis of the
SSS phase because it has not ended at the time of writing. During
the third campaign we also obtained dense ground-based optical
monitoring observations in the $V$-band.

\subsection{Optical observations}
\label{vbandsect}

Ground-based photometric time series of V458\,Vul were obtained
during the period July 8 to August 28 using four different 0.28--0.4\,m
telescopes located in Wales, England, Belgium and New Hampshire, US,
and are equipped with SBIG or Starlight Express cameras. Images
were processed in a standard fashion using MaxIm\,DL and
AIP4WIN, and differential
magnitudes were measured using USNO-B1.0\,1108-0459444,
1109-0453942, and 1108-0459876 as comparison stars. Instrumental magnitudes were then
converted to a pseudo-$V$ band using the AAVSO photometric
sequence for V458\,Vul.

\subsection{\swift\ UV and X-ray observations} 

In the left part of Table~\ref{tab1} we list all \swift\
observations, giving
the date and time when each observation started and ended, the
corresponding number of days after outburst (averaged between
start- and stop times), the ObsID, and the exposure time. Due
to its low-Earth orbit, \swift\ can only observe a given target for
a maximum of $\sim 2$\,ks every 96-minute revolution, and each
observation consists of a number of these short snapshots. The XRT
observations are integrated over all snapshots, while the UVOT
filters can be switched during an observation. The bandpasses for
each filter are given in terms of central wavelengths and full
width at half maximum (FWHM) in the header of the right part
of Table~\ref{tab1}.

\begin{table*}[!ht]
\begin{flushleft}
\renewcommand{\arraystretch}{1.1}
\caption{\label{tab1}Observation log (left) and results (right)}
\begin{tabular}{llllr||ccccccc}
Start Date & Stop Date & Day$^{a}$& \multicolumn{2}{l||}{ObsID \hfill net exp.} & $CR$ (cts/ks)$^b$ & $HR^c$ & uvw1$^d$ & uvw2$^d$ & uvm2$^d$ & u$^d$ \\
&&&&(s)&0.25-10\,keV & & 2600\,\AA$^e$ & 1928\,\AA$^e$ & 2246\,\AA$^e$ & 3465\,\AA$^e$\\
&&&&&& & 693\,\AA$^f$ & 657\,\AA$^f$ & 498\,\AA$^f$ & 785\,\AA$^f$\\
\hline
2007, Aug 09, 16:27 & Aug 09, 16:43&1.19&00030980001&985.6&
 $<3.1$ & -- & -- & $\mathrm{203.00}$ & $\mathrm{114.00}$ & -- \\
\hline
2007, Oct 18, 01:06 & Oct 18, 23:47&71.01&00030980002&6515.4&
 $31.3\,\pm\,2.5$ & +$0.15\,\pm\,0.13$ & -- & -- & -- & -- \\
2007, Nov 01, 01:10 & Nov 01, 02:42&84.57&00030980003&908.1&
 $48.8\,\pm\,8.3$ & +$0.22\,\pm\,0.30$ & -- & $\mathrm{29.40}$ & -- & -- \\
2007, Nov 08, 16:06 & Nov 08, 18:02&92.21&00030980004&2276.7&
 $59.0\,\pm\,5.6$ & +$0.08\,\pm\,0.15$ & -- & -- & -- & $\mathrm{0.04}$ \\
2007, Nov 15, 03:56 & Nov 15, 05:49&98.70&00030980005&1967.4&
 $67.0\,\pm\,7.4$ & +$0.11\,\pm\,0.18$ & $\mathrm{29.00}$ & -- & -- & -- \\
2007, Nov 22, 01:13 & Nov 22, 03:07&105.58&00030980006&1710.4&
 $52.5\,\pm\,6.2$ & $-0.21\,\pm\,0.13$ & -- & -- & $\mathrm{7.17}$ & -- \\
2007, Nov 29, 21:11 & Nov 29, 23:12&113.42&00030980007&2861.3&
 $57.2\,\pm\,4.9$ & $-0.1\,\pm\,0.12$ & -- & $\mathrm{23.20}$ & -- & -- \\
2007, Dec 06, 02:46 & Dec 06, 20:23&119.98&00030980008&2237.1&
 $62.8\,\pm\,5.8$ & $-0.1\,\pm\,0.12$ & -- & -- & -- & $\mathrm{0.06}$ \\
2007, Dec 13, 19:28 & Dec 13, 22:60&127.38&00030980009&1919.7&
 $57.4\,\pm\,6.2$ & +$0.12\,\pm\,0.17$ & $\mathrm{11.60}$ & -- & -- & -- \\
2007, Dec 20, 12:17 & Dec 20, 17:22&134.11&00030980010&2200.0&
 $70.9\,\pm\,7.1$ & $-0.\,\pm\,0.14$ & -- & -- & $\mathrm{6.19}$ & -- \\
2007, Dec 27, 00:02 & Dec 27, 01:59&140.54&00030980012&2386.7&
 $35.9\,\pm\,5.5$ & $-0.15\,\pm\,0.18$ & -- & $\mathrm{74.30}$ & -- & -- \\
\hline
2008, Jun 18, 00:48 & Jun 18, 23:31&315.00&00030980013&7661.3&
 $24.0\,\pm\,1.9$ & $-0.48\,\pm\,0.06$ & $\mathrm{2.47}$ & -- & -- & -- \\
2008, Jul 15, 21:28 & Jul 15, 23:25&342.43&00030980014&2647.1&
 $22.5\,\pm\,3.4$ & +$0.02\,\pm\,0.21$ & $\mathrm{3.20}$ & -- & -- & -- \\
2008, Jul 22, 14:20 & Jul 22, 17:33&349.16&00030980015&1748.6&
 $13.0\,\pm\,3.3$ & $-0.1\,\pm\,0.34$ & $\mathrm{7.76}$ & -- & -- & -- \\
2008, Jul 29, 18:11 & Jul 29, 23:01&356.36&00030980016&3230.7&
 $12.7\,\pm\,2.5$ & $-0.35\,\pm\,0.17$ & $\mathrm{6.69}$ & -- & -- & -- \\
2008, Aug 12, 04:32 & Aug 12, 14:28&369.89&00030980018&4163.0&
 $18.5\,\pm\,2.4$ & +$0.21\,\pm\,0.22$ & $\mathrm{4.22}$ & -- & -- & -- \\
2008, Aug 19, 07:06 & Aug 19, 09:03&376.83&00030980019&2112.7&
 $28.4\,\pm\,4.8$ & $-0.41\,\pm\,0.10$ & $\mathrm{2.45}$ & -- & -- & -- \\
2008, Aug 26, 01:03 & Aug 26, 02:59&383.58&00030980020&1915.4&
 $19.1\,\pm\,3.7$ & +$0.40\,\pm\,0.38$ & $\mathrm{3.03}$ & -- & -- & -- \\
2008, Sep 01, 06:30 & Sep 01, 16:19&389.97&00030980021&4622.5&
 $17.7\,\pm\,2.2$ & $-0.43\,\pm\,0.10$ & $\mathrm{2.29}$ & -- & -- & -- \\
\hline
2008, Sep 09, 03:01 & Sep 09, 14:24&397.86&00030980022&3074.2&
 $45.2\,\pm\,4.2$ & $-0.84\,\pm\,0.02$ & $\mathrm{1.75}$ & -- & -- & -- \\
2008, Sep 14, 01:51 & Sep 15, 22:56&402.57&00030980023&6954.9&
 $175.0\,\pm\,5.2$ & $-0.95\,\pm\,0.00$ & $\mathrm{1.58}$ & -- & -- & -- \\
2008, Sep 16, 08:11 & Sep 16, 19:44&404.84&00030980024&4155.5&
 $207.3\,\pm\,7.4$ & $-0.97\,\pm\,0.00$ & $\mathrm{1.63}$ & -- & -- & -- \\
2008, Sep 17, 00:02 & Sep 17, 22:59&405.50&00030980026&12119.3&
 $408.2\,\pm\,5.9$ & $-0.97\,\pm\,0.00$ & $\mathrm{1.33}$ & -- & -- & -- \\
2008, Sep 18, 01:54 & Sep 18, 21:41&406.57&00030980027&10581.0&
 $180.8\,\pm\,4.3$ & $-0.95\,\pm\,0.00$ & $\mathrm{1.56}$ & -- & -- & -- \\
2008, Sep 21, 11:42 & Sep 21, 13:40&409.98&00030980030&2095.2&
 $702.3\,\pm\,18.6$ & $-0.98\,\pm\,0.00$ & $\mathrm{0.99}$ & -- & -- & -- \\
2008, Sep 23, 00:51 & Sep 23, 16:49&411.53&00030980032&1860.5&
 $631.5\,\pm\,19.8$ & $-0.99\,\pm\,0.00$ & $\mathrm{1.15}$ & -- & -- & -- \\
2008, Sep 25, 12:01 & Sep 25, 17:04&413.99&00030980033&2222.6&
 $282.0\,\pm\,11.5$ & $-0.99\,\pm\,0.00$ & $\mathrm{1.16}$ & -- & -- & -- \\
2008, Sep 30, 03:18 & Sep 30, 19:35&418.63&00030980034&1905.4&
 $584.9\,\pm\,17.8$ & $-0.99\,\pm\,0.00$ & $\mathrm{1.08}$ & -- & -- & -- \\
2008, Oct 01, 06:10 & Oct 01, 11:08&419.75&00030980035&2150.2&
 $675.8\,\pm\,18.0$ & $-0.98\,\pm\,0.00$ & $\mathrm{1.23}$ & -- & -- & -- \\
2008, Oct 02, 09:28 & Oct 02, 16:13&420.89&00030980036&2020.3&
 $702.6\,\pm\,18.7$ & $-0.98\,\pm\,0.00$ & $\mathrm{1.07}$ & -- & -- & -- \\
2008, Oct 03, 00:17 & Oct 03, 03:42&421.51&00030980037&1825.5&
 $762.6\,\pm\,22.7$ & $-0.97\,\pm\,0.00$ & $\mathrm{0.98}$ & -- & -- & -- \\
\hline
\end{tabular}

$^{a}$After outburst (2007, Aug. 8.54), average between start- and stop times\ $\bullet$ $^b$Corrected for PSF losses and bad columns\\
$^c$Hardness ratio $HR=(H-S)/(H+S)$ with $S=0.25-1$\,keV and
$H=1-10$\,keV\ $\bullet$
$^d$fluxes in $10^{-15}$\,erg\,cm$^{-2}$\,s$^{-1}$
$^e$Central wavelength \citep{poole07}\ $\bullet$ $^f$Band width \citep[FWHM;][]{poole07}

\renewcommand{\arraystretch}{1}
\end{flushleft}
\end{table*}

\subsubsection{Reduction procedures}

We have used the \swift\ tools version 2.9 with CALDB 2.9, which are
part of the HEADAS software
package\footnote{http://swift.gsfc.nasa.gov/docs/software/lheasoft/}
to produce level 2 products (sky images for UVOT observations and
cleaned event files for XRT observations).
UV fluxes were extracted from the sky images using standard
tools, and X-ray count rates were obtained from the cleaned event files
following the procedure described by \cite{swnovae}.

We have determined the source count rate using a
circular source extraction region with radius 10 pixels
($=23.6\arcsec$), and an annular background region with inner and
outer radii of 10 and 80 pixels, respectively. This outer radius
($=188.8\arcsec$) was chosen so as to exclude a nearby
X-ray source.
The analysis procedure adopted from \cite{swnovae} corrects for
source photons contained in the wings of the point spread
function (PSF) that are recorded in the background region.
We have
also corrected for losses in cases where the source position
coincided with bad columns on the detector. The bad columns
can be identified on the exposure maps, and we have calculated a
correction factor by comparing the effective areas extracted from the
level 2 event file and from the exposure map.

\begin{figure}[!ht]
\resizebox{\hsize}{!}{\includegraphics{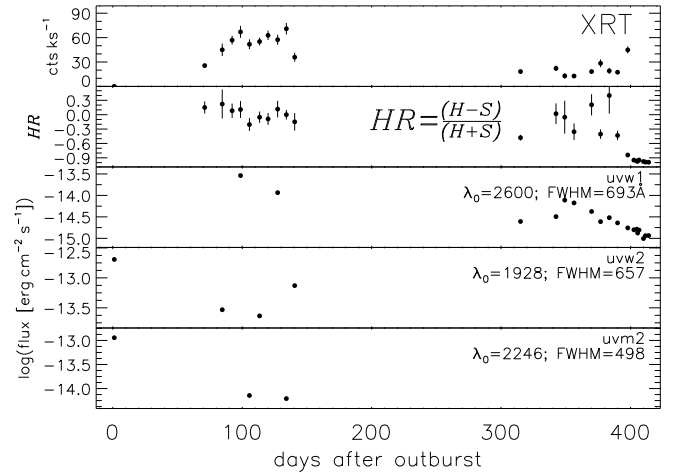}}
\caption{\label{uvot}Comparison of the X-ray light curve, X-ray hardness
(top two panels; with $S$ and $H$ the count rates in the bands $0.25-1$\,keV
and $1-10$\,keV, respectively), and UV fluxes.
After day 400 the X-ray count rate is higher
than 90 counts per ks due to the SSS phase component, and these rates
don't fit in the top graph.
}
\end{figure}

In the right part of Table~\ref{tab1}, we list the corrected XRT
count rates with 1-$\sigma$ uncertainties or 95-\% upper limits,
hardness ratios, and UVOT fluxes. The XRT hardness ratio is defined
conventionally here as $HR=(H-S)/(H+S)$, with $H$ and $S$ the
number of counts in a hard and soft X-ray band, respectively.
With this definition, the hardness ratio is +1 for very hard
sources and -1 for
very soft sources. We have calculated the hardness ratios using an
energy band of $0.25-1$\,keV for $S$ and $1-10$\,keV for $H$.
We chose
these ranges to yield a roughly equal number of counts in each band
in order to minimize the uncertainties in the hardness ratios.

\subsubsection{Four observing campaigns}

 In Fig.~\ref{uvot} we illustrate the evolution of X-ray count
rates and hardness ratios (top two panels) and the UVOT fluxes
(in units of $10^{-15}$\,erg\,cm$^{-2}$\,s$^{-1}$, bottom three
panels) during the first $\sim 420$ days.

 For day 1.2, we find a 95-per cent upper limit of 0.0021\,cps.
From the spectral models presented in \S\ref{xspecsect} and
Table~\ref{models}, and assuming a distance of 13\,kpc, this
implies an X-ray luminosity of less than
$10^{33}$\,erg\,s$^{-1}$ for the XRT band of $0.25-10$\,keV.
The uvw2 flux corresponds to a luminosity of
$4\times 10^{33}$\,erg\,s$^{-1}$ in the waveband $1271-2585$\,\AA.

 From day 71 to day $\sim 100$, the X-ray count rate increases
monotonically, while the hardness stays about the same. A small
drop in count rate after day $\sim 100$ is accompanied by a
reduction in hardness. The spectrum does not evolve into that
of an SSS at this time. We have studied the likelihood for the
hardness ratio being variable during campaign II.
All values are consistent with their mean, and assuming a
constant ratio of $+0.02$ yields a reduced $\chi^2=0.8$.
The UV flux has significantly decreased from campaign I to II,
and the corresponding UV luminosity
varies between $5\times 10^{32}$\,erg\,s$^{-1}$ and
$1.5\times 10^{33}$\,erg\,s$^{-1}$ in the waveband
$1271-2585$\,\AA, thus a factor around 10 lower than in campaign I.

Campaign III starts with a lower X-ray count rate compared to
campaign II, but the spectrum is significantly softer.
However, 30 days into campaign III, the hardness ratio was back up
again, and the observation taken on day 315 is the only campaign-III
observation that has a SSS component. In campaign III we have
used the same UVOT filter (uvw1) for all observations, and the
UV luminosity varies between $6\times 10^{31}$\,erg\,s$^{-1}$ and
$1.6\times 10^{32}$\,erg\,s$^{-1}$ in the waveband
$1900-3300$\,\AA.

\begin{figure}[!ht]
\resizebox{\hsize}{!}{\includegraphics{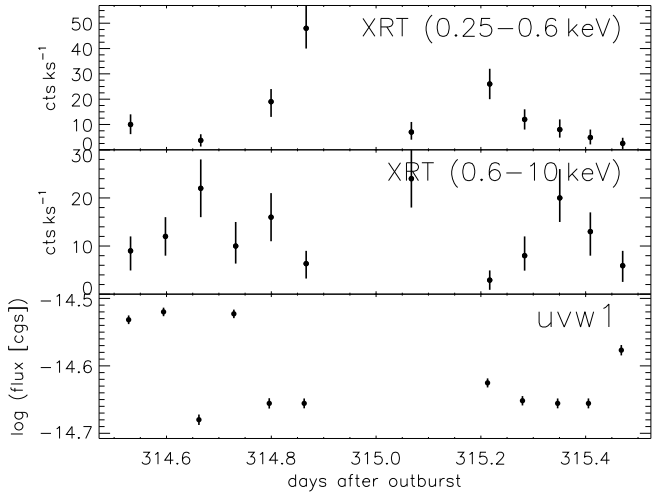}}
\caption{\label{lc}X-ray soft and hard light curves and uvw1 light
curve around day 315. A small flare is seen in the soft X-ray
component. The hard X-ray count rate and the uvw1 flux are suggestive
of an anticorrelation, but a Spearman rank test with eleven values
gives a coefficient of $-0.43$, which is not significant.}
\end{figure}

In Fig.~\ref{lc} we present details of the first observation of
campaign III (day 315), showing detailed X-ray light curves in two
energy bands and the simultaneous uvw1 light curve. Three data points
are missing because of tracking problems in orbits 7, 8, and 10,
during which the source was not in the field. We have excluded these
orbits for extracting the average count rate listed in
Table~\ref{tab1}. In the soft band a flare-like event with a peak
around day 314.86 and a duration less than 0.4 days is seen which
has no counterpart in harder X-rays or the uvw1 fluxes. A similar
flare event has been reported by \cite{drake03} in the nova
V1494\,Aql (see also \citealt{rohrbach09}). The hard
X-ray count rate seems to be anticorrelated with the uvw1 fluxes,
as each increase in X-ray count rate coincides with a decrease in
UV brightness. In order to determine whether there was a
correlation between the X-ray count-rate and the UV fluxes, we
have computed a Spearman rank coefficient\footnote{Using
the web tool Wessa, P. (2008), Free Statistics Software, Office for
Research Development and Education, version 1.1.23-r2, URL
http://www.wessa.net/} of $r_{\rm S}=-0.43$. 
This corresponds to a $<2\sigma$ probability of an
anti-correlation which thus cannot be considered statistically
significant.

\begin{figure}[!ht]
\resizebox{\hsize}{!}{\includegraphics{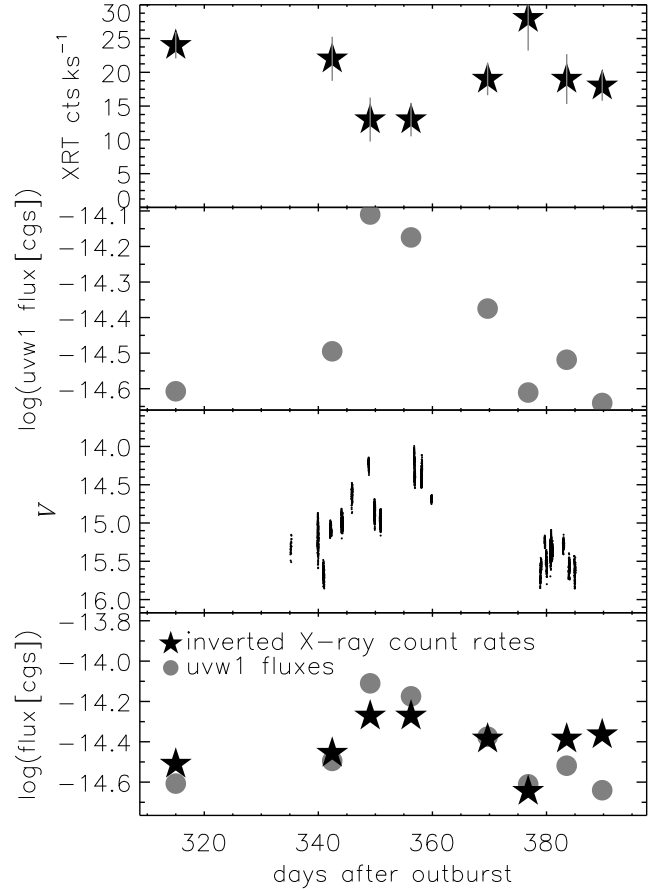}}
\caption{\label{anticorr}From top to bottom: XRT count rates,
uvw1 fluxes, $V$-band magnitudes, and comparison of
X-ray and uvw1 fluxes for the observations taken between
day 315 and 390. The X-ray count rates are extracted from the
hard band ($0.6-10$\,keV). We have converted the
X-ray count rates to scale with the uvw1 fluxes. The comparison
of the inverted X-ray light curve with the uvw1 light curve
is shown in the bottom panel and illustrates the anticorrelation.
The Spearman rank test on the first seven datapoints yields a
97-\% probability that the anticorrelation is real, but with
the last data point (day 390) included, the indication for an
anticorrelation is much weaker. A reason could be that at that
time the UV flux is affected by the emerging SSS component (see text).
}
\end{figure}

 In Fig.~\ref{anticorr} we show the X-ray and uvw1 datapoints
for all observations taken between days 315 and 390 (campaign III).
In order to study only the hard emission, we extracted the
X-ray count rates in the band pass $0.6-10$\,keV. Our choice
of energy range here is based on the spectrum that is shown
in Fig.~\ref{xrtspec_2} (dark grey shade), where it can be seen that
below 0.6\,keV the emission is dominated by the new soft component.
For comparison
we include the $V$-band magnitudes described in \S\ref{vbandsect},
and they show the same long-term variations as the uvw1 data.

\begin{figure}[!ht]
\resizebox{\hsize}{!}{\includegraphics{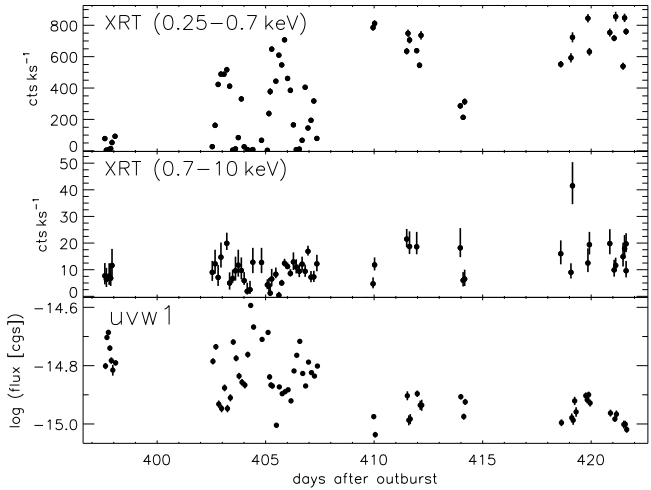}}
\caption{\label{lc_sss}Soft and hard X-ray light curves (top
two panels) and corresponding uvw1 fluxes for campaign IV.
The pivot energy is chosen higher to avoid contamination of
the hard count rates by SSS emission. All X-ray count rates
are corrected for pile up (see text).
}
\end{figure}

 From the top two panels it is again suggestive that the
hard X-ray count rates anticorrelate with the uvw1 fluxes.
In order to illustrate this anticorrelation, we have rescaled
the X-ray count rates to the uvw1 fluxes and inverted the X-ray
light curve. In the bottom panel of Fig.~\ref{anticorr} we show
the inverse X-ray light curve in comparison to the uvw1 fluxes.
The same up- and down trends can be seen. The Spearman rank
coefficient is $r_{\rm S}=-0.59$ for all eight datapoints
but $r_{\rm S}=-0.93$ if excluding the observation taken on
day 390. A reason to exclude this data point would be that
the UV flux in this observation could already be affected
by emission from the WD, as a bright SSS spectrum was observed
only a week
later (see Fig.~\ref{lc_sss}). The latter correlation yields a 97-\%
probability, which is suggestive of a real anticorrelation.

 Next, we have tested the proposed orbital solution given by
\cite{v458orbit}, $max=2454461.479 + 0.58946^{\rm days}\times
E$, with $max$ being the barycentric Julian Day of optical
maximum. We have converted the arrival times of the photons from all
observations taken during campaign III as well as the observing
times of the optical $V$-band magnitudes (see \S\ref{vbandsect}),
but in none of the bandpasses do we see
any evidence for modulation at this period. We performed
additional period checks on the uvw1 light curve of days
315 to \dend\ but found no convincing evidence for any
periodic changes. We note that a period search is
difficult with the interrupted light curves available
from \swift.

 After day 397 the SSS phase started, and a campaign of
monitoring at higher cadence was initiated (campaign IV;
see \citealt{v458sss}). In this paper we focus on the
evolution of the hard component and postpone the discussion
of the SSS phase to a future paper. In order to trace the
evolution of the hard component separately from the SSS
emission, we have extracted the light curves in the energy
bands $0.25-0.7$\,keV and $0.7-10$\,keV, respectively.
We chose a higher pivot energy of 0.7\,keV in order to
exclude any SSS emission from the hard band pass (see
Fig.~\ref{xrtspec_2}). The
brightness of the SSS component leads to significant
effects of pile up, whereby two incoming X-rays are
counted as one count of twice the energy. In order to
avoid contamination of the hard component by these
counts, we have corrected for pile up by excluding
the central 5 pixels for the time intervals when the
total count rate was between 0.3 and 0.5 cps, and 7 pixels
when it was above 0.5 cps. The extracted count rates
have then been upscaled, yielding the PSF-integrated
total number of counts. We note, however, that the pile
up correction leads to small raw numbers of counts, which
leads to large statistical uncertainties. Detailed
analyzes are thus not possible for the faint hard
component. In Fig.~\ref{lc_sss}, we compare
the evolution of the corrected soft ($0.25-0.7$\,keV) and hard
($0.7-10$\,keV) X-ray count rates and of the uvw1 fluxes
(top to bottom, respectively). The hard component is still
present, and the corrected count rates are consistent with those
found in the pre-SSS observations. In the middle panel of
Fig.~\ref{lc_sss} one can see that the hard count rate exhibits
variability that is uncorrelated to that of the soft flux (top
panel).
The hard component is thus unaffected by the SSS emission and
can be assumed to continue to fade at a very slow rate. However,
no significant reduction in count rate can be identified from
the middle panel of Fig.~\ref{lc_sss}. 

\begin{figure}[!ht]
\resizebox{\hsize}{!}{\includegraphics{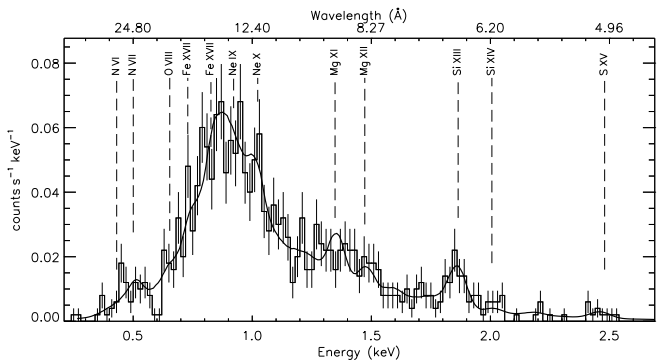}}
\caption{\label{xrtspec}XRT spectrum integrated over days 71-140.5
(total of
25 ks). The solid line is a thermal model (VAPEC with variable
abundances) with the parameters listed in Table~\ref{models}.
Labels are included at energies where strong lines are expected.
The instrumental response at 1\,keV is 200\,eV FWHM, and strong
isolated lines can be identified.
}
\end{figure}

\subsection{Spectral analysis}
\label{xspecsect}

In Fig.~\ref{xrtspec} we show the X-ray spectrum integrated over
the time interval from day 71 to 140.5 (campaign II). We have
included labels at the energies where strong emission lines can
in principle arise. Strong emission peaks appear at the wavelengths
of silicon and nitrogen lines, while the lines of magnesium are
weaker. The full width at half maximum (FWHM) of the XRT spectral
response is $\sim 200$\,eV at 1\,keV.
Between $0.7-1.0$\,keV the H-like and He-like Ne lines overlap with
strong Fe\,{\sc xvii} L-shell lines. The combined XRT spectrum
resembles the \suzaku\ spectrum taken on day 88 after outburst
\citep[see][]{masahiro08}.

We have modeled the XRT spectrum using an isothermal VAPEC model
\citep{smith01}, which represents a collisional, optically thin
plasma. We have used the {\sc xspec} fitting package, version
11.3.2ag \citep{xspec}, allowing
variable abundances for N, O, Ne, Mg, Si, S, and
Fe, with the abundances of all other elements fixed at the solar
values of \citet{agrev89}. In order to account for absorption in
the interstellar medium (ISM) we have used the
{\sc tbabs} module with the single parameter $N_{\rm H}$, the
neutral hydrogen column density in cm$^{-2}$, assuming solar
composition of the ISM \citep{wilms00}. We have optimized the
parameters using {\sc cstat} statistics with which we avoid
having to rebin the spectrum.
We have used the {\sc error} command to compute 90\%
uncertainty ranges.

The results and 90\% uncertainty ranges for each parameter are listed
in Table~\ref{models}. The temperature, k$T$, neutral hydrogen column
density, $N_{\rm H}$, element abundances, and X-ray luminosity are in
good agreement with the parameters found from \suzaku\ spectra by
\cite{masahiro08}, indicating little evolution in the hot plasma
between the respective epochs of the \swift\ average spectrum
analyzed here (days 71-141) and the \suzaku\ (day 88) spectra.
From the average count rate we determine a rough count rate-to-luminosity
conversion factor of $5\times10^{35}$\,erg\,s$^{-1}$\,cps$^{-1}$,
thus the upper limit of 0.002\,cps for day 1.2 converts to an
upper limit to the luminosity of $10^{33}$\,erg\,s$^{-1}$ over
$0.25 - 10$\,keV.
Element abundances are all consistent with solar values, except for
that of Fe, which we find to be at least a factor of two lower.
Underabundance of Fe is possibly common in post-AGB stars which
have undergone dredge-up of s-processed material.
Low Fe abundances have also been found in the X-ray spectra of
novae, e.g., RS\,Oph \citep{rsophshock} and V382\,Vel
\citep{ness_vel}.
Unfortunately, abundances of N and O could not be usefully
constrained. The upper limit found for N is relatively high,
and allows for the possibility that the hot gas is enriched by
CNO processing.
The plasma temperature lies in the range 0.58-0.7\,keV.

The best-fit model is included with a solid line in Fig.~\ref{xrtspec}.
The model represents a satisfactory fit. To give a goodness criterion,
we have calculated a value of reduced $\chi^2=0.85$ (263 degrees of
freedom), using the errors extracted with the spectrum in the original
binning.
In addition we have carried out the {\sc goodness} command with 10,000
trials, and found that 41\% of realizations have a fit statistic
better than the best fit. For comparison, the ''ideal'' fit yields 50\%
of all realizations in the Monte Carlo calculation to be better, i.e.,
our fit is formally acceptable.

\begin{table}
\begin{flushleft}
\renewcommand{\arraystretch}{1.1}
\caption{\label{models}Spectral Models to XRT spectra of days 71-140.5 and 315}
\begin{tabular}{llr}
\hline
Param. & Unit & Value$^a$\\
\hline
\multicolumn{3}{l}{VAPEC model to Campaign II, combined XRT spectrum}\\
k$T$ &keV\dotfill & $0.64\,\pm\,0.06$\\
$N_{\rm H}$ &$10^{21}$cm$^{-2}$\dotfill & $2.7\,\pm\,0.9$\\
A(N)   & solar$^b$\dotfill     & $<27.9$\\
A(O)   & solar$^b$\dotfill     & $<1.1$\\
A(Ne)  & solar$^b$\dotfill     & $<0.9$\\
A(Mg)  & solar$^b$\dotfill     & $0.5\,\pm\,0.3$\\
A(Si)  & solar$^b$\dotfill     & $0.8^{+1.0}_{-0.4}$\\
A(S)   & solar$^b$\dotfill     & $0.7^{+1.0}_{-0.5}$\\
A(Fe)  & solar$^b$\dotfill     & $0.2^{+0.3}_{-0.1}$\\
flux$^c$ &$10^{-12}$\,erg\,cm$^{-2}$\,s$^{-1}$\dotfill & $1.15\,\pm\,0.65$\\
$L_X^{c,d}$     & $10^{34}$\,erg\,s$^{-1}$\dotfill  & $2.3\,\pm\,1.3$\\
$\chi^2_{\rm red}$ ($dof$) &\dotfill & 0.85 (263)\\
\hline
\multicolumn{3}{l}{Blackbody (bb) plus VAPEC fit to day 315 spectrum}\\
$N_{\rm H}$ &$10^{21}$cm$^{-2}$\dotfill & $3.1\,\pm\,0.4$\\
$T_{\rm eff}$(bb) & eV\dotfill & $22.8^{+9.4}_{-5.1}$\\
$\log(L_{\rm bol})^d$ & erg\,s$^{-1}$\dotfill & $39\,\pm\,2$\\
radius$^d$ & $10^3$\,km\dotfill & $170^{+2640}_{-161}$\\
k$T$(VAPEC) & eV\dotfill & 0.55\\
flux$^c$(VAPEC) &$10^{-12}$\,erg\,cm$^{-2}$\,s$^{-1}$\dotfill & $0.30\,\pm\,0.15$\\
\hline
\end{tabular}

$^a$90\% uncertainty ranges\ $\bullet$ $^b$\cite{agrev89}\\
$^c$ Unabsorbed over range $0.25-10$\,keV\\
$^d$assuming distance 13\,kpc
\renewcommand{\arraystretch}{1}
\end{flushleft}
\end{table}

\begin{figure}[!ht]
\resizebox{\hsize}{!}{\includegraphics{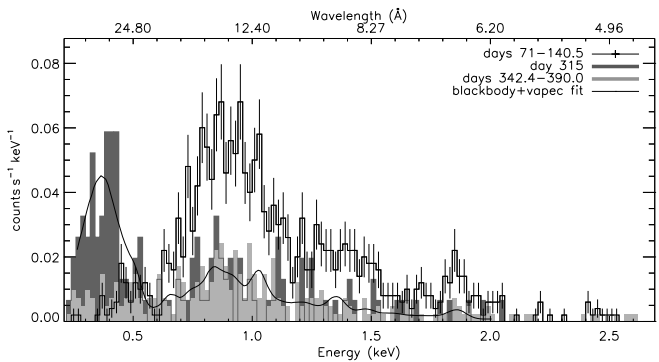}}
\caption{\label{xrtspec_2}XRT spectrum from Fig.~\ref{xrtspec} in
comparison to the observation taken on day 315 (dark gray shading)
and the sum of observations taken between days 342.4 and \dend\
(light gray shading).
Overplotted is a model consisting of a blackbody and a VAPEC
with parameters given in Table~\ref{models}.
}
\end{figure}

 In Fig.~\ref{xrtspec_2} we compare the sum of all campaign-II
XRT spectra (solid line, see Fig.~\ref{xrtspec}), the observation taken
on the first day of campaign III (day 315 in dark shading),
and the sum of the last seven campaign-III observations (days
342.4-\dend\ in light shading). From day 140.5 to 315, the hard component
between 0.7 and 1.0\,keV has faded, but below 0.5\,keV a significantly
higher emission level was observed on day 315. This soft component leads to the lower
hardness ratio listed in Table~\ref{tab1}. After day 342.4 the
hardness ratio is higher again, and the soft component is not
present in the combined XRT spectrum from days 342.4-\dend.
However, the hard component has not changed from day 315
to the summed observation of days 342.4-\dend.

We have fitted this soft component with a blackbody.
In order to constrain the blackbody fit against
arbitrarily high temperatures, we have added a VAPEC model
with the abundances fixed at the same values as those listed in
Table~\ref{models}. The best-fit blackbody plus VAPEC model is shown
with a thin
solid line in Fig.~\ref{xrtspec_2}, and the results are given
in the lower part of Table~\ref{models}. The blackbody
temperature is reminiscent of a typical SSS spectrum and the
value of $N_{\rm H}$ is consistent with that found from
campaign II. The luminosity range encompasses the Eddington
luminosity of a 1M$_\odot$ white dwarf, but it is not well
constrained owing to the uncertainty in effective temperature
and $N_{\rm H}$. Similarly, the radius derived from the
temperature and luminosity is very uncertain and is
consistent with an extended white dwarf.

In the spectrum taken on day 315, the hard component above
0.6\,keV is difficult to model owing to lack of signal. 
We find a temperature of the VAPEC component that is
consistent with that found from the sum of the campaign-II
observations (top part of Table~\ref{models}). The flux of the
VAPEC component is a factor 3.8 lower than that seen in
the previous observation on day 315, and the
hard component has thus faded by this amount. The light
gray-shaded spectrum shown in Fig.~\ref{xrtspec_2} indicates
the sum of the XRT spectra over days 342.4-\dend, and there is no
significant difference between the dark- and light shaded spectra
except for the disappearance of the soft component. This indicates
that the hard component evolves slowly.

\section{Discussion}
\label{disc}

The Classical Nova V458\,Vul has been monitored in X-rays and the
ultraviolet between days 1 and 422 after outburst using the \swift\
satellite. No X-ray
emission was detected 1.2 days after outburst, but the source
was bright in the UV. The early optical light curve shows two
strong peaks on days 4 and 10 after outburst with dramatic
changes in line-width velocities \citep{tarasova07}. Since no
X-ray emission was detected three days before the first peak, it
is possible that the peaks were in some way associated with the
trigger for the X-ray emission. However, no X-ray observations
were taken during and immediately after the peaks occurred, and
we can not discuss whether this peculiar behavior is in any way
related to the slowly evolving hard X-ray emission. We thus
encourage systematic monitoring through this evolutionary phase
of future nova outbursts in order that such scenarios can be
confirmed or rejected.

The second monitoring campaign took place between days 71 and
140.5 after outburst, and a variable, hard X-ray source was
observed. We have modelled the combined X-ray spectrum and
found a good fit with an isothermal collision-dominated optically-thin
plasma model. Our model parameters are consistent with the spectral
model fits to \suzaku\ spectra obtained on day 88 after outburst
\citep[see][]{masahiro08}. We find a chemical composition
consistent with a solar mixture, except for Fe which appears
underabundant by a factor of at least 2. In CNO-cycled material,
nitrogen is expected to be overabundant, however, our
data allow us only to derive an upper limit to the N abundance.
This limit is high enough that CNO ashes are not ruled out
from the observations. The low Fe abundance might reflect the
composition of the post-AGB progenitor star of the white dwarf,
and Fe underabundance has been found in several other novae as
well.

The first observation of the third campaign between 315 and \dend\
days after outburst shows the first glimpse of the SSS phase, but the
proper SSS phase does not start until 80 days later \citep{v458sss}.
The soft component ($0.25-0.6$\,keV) is highly variable
with a flare-like event in X-rays, while the hard component
($0.6-10$\,keV) shows a different pattern of variability.

We have searched for the orbital period proposed by
\cite{v458orbit} but found no evidence for it in either 
X-ray, UV, or optical.
Variations seen in the UV are in closer relationship to those
in the harder X-ray component than in the soft band. During the
entire third campaign, the UV brightness is anti-correlated with
the hard X-ray count rate at the 97-\% probability level.
The hard component during campaign III is a factor 3.8 fainter
than during campaign II and can also be modelled with an
optically-thin plasma spectrum. The signal-to-noise does not
allow us to detect any changes in temperature. If the emission
originates from the same plasma as in campaign II, then it has
only cooled slightly, and the hard component fades extremely
slowly. Owing to additional short-term variability within each
observing campaign (see Fig.~\ref{uvot}), we cannot determine
an analytic decay law.

 The fourth campaign is dominated by the SSS emission
\citep{v458sss}. At the time of writing, the SSS phase
has not ended yet, and we postpone the analysis of the
entire SSS phase. Meanwhile, the hard component appears virtually
unaffected, which is consistent with expectations that it
originates in regions far away from the photosphere of the
white dwarf. Our finding of anticorrelations between the
hard component and the UV flux suggests that the UV emission
originates at least partly from those outer regions as
well. However, the fact that the UV emission also correlates
with the SSS emission \citep{v458sss} suggests that we are seeing a mixture
of different UV sources. This may explain why the detection
of the anticorrelation between X-ray and UV emission during
campaign III is so difficult.

While the SSS phase is the most prominent evolutionary phase
in X-rays, additional X-ray production mechanisms are poorly
understood. While we are not able to derive a decay law,
it is clear that the hard X-ray component fades over a
time scale of more than a year and is still present after
the SSS phase started. Since the density in the ejecta must
have decreased for the SSS spectrum to emerge, X-ray emission
originating from shocks within the ejecta might be expected to
have faded by this time \citep{lloyd92,obrien94}, yielding a
shorter time scale than that observed.
However, it is possible that the emission observed in
campaign II could have its origin in internal shocks as
discussed by \cite{masahiro08}. In that case the hard component
observed during campaigns III and IV would have a different origin.
On the other hand, \cite{orio96} 
state that they can't be definitive either about their 
late-time single epoch detection of Pup 1991 coming from 
internal shocks in the ejecta or from accretion at a high rate.

 Other sources of the hard X-ray emission could have been shock
interactions of the expanding ejecta with the stellar wind of the
companion or with the surrounding medium which includes the
innermost regions of the planetary nebula as is suggested for
GK\,Per from direct imaging (see \citealt{bode04} and
references therein;
and also \citealt{balman_gkper} and \citealt{vrielmann05}).
 Shocks with the stellar wind of the companion
occur most notably in the case of long-period symbiotic novae
in which the mass-donor star is an evolved object
with a massive wind \citep[e.g., RS\,Oph:][]{bode06}.
 However, \citet{Wessonetal08}
note that the pre-outburst colors are consistent with
an O-type spectral class, and attribute this to either the
post-AGB evolution of the PN progenitor, or to an accretion
disk. It is possible that the more tenuous circumstellar
environment of this system still provides for significant
shock-induced X-rays. The timescale for dissipation of the
explosion energy in such a tenuous medium would also
be much longer than in the denser wind of the symbiotic case.

\section{Summary and Conclusions}
\label{concl}

 The four observing campaigns can be summarized as follows:
\begin{itemize}
\item The early evolution, before day 70 after outburst, showed
an erratic optical light curve \citep{tarasova07}, but we have
only one X-ray observation on day 1.2 that yields a non-detection.
This non-detection suggests that the hard X-ray component has not
started its evolution with the outburst directly. The two peaks in
the optical light curve on days 4 and 10 after outburst could be
related to the trigger for the X-ray emission. For example, X-ray
and optical emission could have been produced on impact of the
expanding ejecta with denser regions of the ambient medium,
possible related to the proposed planetary nebula. However, in
order to test such scenarios, denser coverage in X-rays during
the early evolution is needed.
\item The first X-ray detection on day 71 after outburst is not
an SSS spectrum, but a hard component that continues to increase
until $\sim $day 100 after outburst. The X-ray spectrum is consistent
with collisional plasma, yielding typical nova abundances. The X-ray
light curve is variable during the second campaign from day 71-140.5.
\item The third campaign, starting on day 315 after outburst, shows
a first appearance of a SSS component, but this component disappears
again, and is not observed again until day 397 after outburst. We
see a suggestion of an anticorrelation between the hard component and
the UV emission, which would indicate that the UV emission may at least
partially originate from the same outer regions as the hard X-ray
emission.
\item The fourth campaign contains a bright SSS component that
is not analyzed in this paper. The hard X-ray component is still
present, suggesting that it is unaffected by the SSS component, and
that it fades extremely slowly. If the UV flux is anticorrelated to
the SSS component and to the hard X-ray component (see above)
then the observed UV flux originates from
the photosphere around the white dwarf and from the outer regions
at the same time.
\end{itemize}

 We have discussed several production mechanisms for the hard X-ray
component. For internal shocks within the ejecta we expect a shorter
decay time than observed \citep{lloyd92,obrien94}, however,
it is possible that the earlier hard X-ray emission (before day 140)
could have originated from within the ejecta \citep{masahiro08}.
Detectable emission from shocks arising from ejecta interaction with
a stellar wind would require a red giant secondary as in symbiotic
novae. If this nova occurred within a planetary nebula as suggested
by \cite{Wessonetal08}, then the interaction of the nova ejecta with 
the innermost regions of material associated with the ejection of the 
PN may be the most plausible source of the long-lasting hard X-ray 
emission.

\acknowledgments

We thank the \swift\ PI Neil Gehrels and
\swift\ schedulers for their support of these TOO observations.
J.-U.N. and M.T. gratefully acknowledge support provided by NASA
through \chandra\ Postdoctoral Fellowship grants PF5-60039 and
PF6-70044, respectively, awarded by the \chandra\ X-ray Center, which
is operated by the Smithsonian Astrophysical Observatory for NASA
under contract NAS8-03060. JJD was supported by NASA contract NAS8-39073
to the {\em Chandra X-ray Center} during the course of this research.
D.T. acknowledges support from a JSPS fellowship. S.S. received
partial support from NSF and NASA grants to ASU. K.L.P., A.P.B,
P.A.E., and J.P.O. acknowledge support from STFC.
D.S. acknowledges a STFC Advanced Fellowship.

\bibliographystyle{apj}
\bibliography{cn,astron,jn,rsoph}

\end{document}